\UseRawInputEncoding
\documentclass[aps,onecolumn,superscriptaddress,groupedaddress,nofootinbib]{revtex4}
\usepackage{graphicx}
\usepackage{dcolumn}
\usepackage{bm}
\usepackage{mathrsfs}
\usepackage{hyperref}
\usepackage{amsmath}
\usepackage{amssymb}
\usepackage{bm}
\usepackage{color}
\usepackage{float}
\usepackage{dcolumn}
\usepackage{multirow}
\usepackage{changepage}
\usepackage{enumerate}
\hypersetup{pdftex,colorlinks=true,linkcolor=blue,citecolor=red,menucolor=black,urlcolor=blue,filecolor=blue}

\newcommand{\mev}{\textrm{ MeV}}
\newcommand{\gev}{\textrm{ GeV}}

\begin{document}

\title{Repercussion  of the $a_0(1710)$ [$a_0(1817)$] resonance and future developments}
\date{\today}

\author{E.~Oset}
\email{oset@ific.uv.es}
\affiliation{Departamento de F\'{\i}sica Te\'orica and IFIC, Centro Mixto Universidad de
Valencia-CSIC Institutos de Investigaci\'on de Paterna, Aptdo.22085,
46071 Valencia, Spain}

\author{L.~R. Dai}
\email{dailianrong@zjhu.edu.cn}
\affiliation{School of Science, Huzhou University, Huzhou 313000, Zhejiang, China}

\author{L. S. Geng}
~\email{lisheng.geng@buaa.edu.cn}
\affiliation{School of Physics, Beihang University, Beijing 102206, China}


\maketitle


In a recent paper the BESIII Collaboration reported the observation of a scalar meson of spin-parity $J^P=0^+$ with isospin $I=1$, branded as $a_0(1817)$. The state is seen as a peak in the $K_S^0  K^+$ mass distribution in the $D_s^+ \to  K_S^0 K^+ \pi^0$ decay \cite{beschar}. Its mass and width are reported as
\begin{eqnarray}\label{eq:a0}
M_{a_0}=1817\pm 8_{\rm stat}\pm 20_{\rm sys}\mev; \quad \Gamma_{a_0}=97 \pm 22_{\rm stat}\pm 15_{\rm sys}\mev
\end{eqnarray}
Prior to this finding, a BABAR experiment reported an $a_0(1710)$ state
from the $\pi^+ \eta$, $\pi^- \eta$ mass distributions in the $\eta_c \to \pi^+ \pi^- \eta$ decay \cite{babar}. The mass and width of this state are obtained as
$M=1709 \pm 5_{\rm stat}\pm 2_{\rm sys}\mev$, $\Gamma=110 \pm 152_{\rm stat}\pm 11_{\rm sys}\mev$.

The new $a_0$ resonance in the sector of light quarks comes as a real surprise at a moment when, however, many new  mesonic states are obtained in the heavy
quark sector \cite{reviewme}. An isospin $I=0$,  $f_0(1710)$ resonance has, however, been known for quite some time \cite{pdg}.
Another BESIII experiment reporting on $D_s^+ \to \pi^+ f_0(1710)$ has brought new elements to this discussion.  Indeed in Ref. \cite{besone} it was
found the branching fraction
\begin{equation}\label{eq:oBr}
 {\mathrm{Br}}[D_s^+ \to  \pi^+ ``f_0(1710)";~``f_0(1710)" \to K^+ K^-]=(1.0 \pm 0.2\pm 0.3)\times 10^{-3} \,,
\end{equation}
while from Ref. \cite{bestwo} it was found that
\begin{equation}\label{eq:nBr}
 {\mathrm{Br}}[D_s^+ \to  \pi^+ ``f_0(1710)";~ ``f_0(1710)" \to K_S^0 K_S^0]=(3.1 \pm 0.3\pm 0.1)\times 10^{-3} \,,
\end{equation}
where $``f_0(1710)"$ was supposed to be the $f_0(1710)$ resonance. Yet, it was concluded that it could not be, because from Eqs.~\eqref{eq:oBr},
\eqref{eq:nBr} one finds
\begin{eqnarray}\label{eq:exR1}
R_1=\frac{\Gamma(D_s^+ \to \pi^+ ``f_0(1710)" \to \pi^+ K^0 \bar{K}^0)}
{\Gamma(D_s^+ \to \pi^+ ``f_0(1710)" \to  \pi^+ K^+ K^-)} =6.20 \pm 0.67  \,.
\end{eqnarray}
If $``f_0(1710)"$ was the $f_0(1710)$ resonance this latter ratio should be $1$. It was concluded that hidden below, or around the $f_0(1710)$,
there should be an $I=1$ resonance responsible for this surprising large ratio. Indeed, since
\begin{eqnarray}\label{eq:wfnK}
|K \bar{K}, I=0\rangle &=& -\frac{1}{\sqrt{2}} \big(K^{0}\bar{K}^{0} \, + \,K^{+} K^{-}  \big) \,,\nonumber  \\
|K \bar{K}, I=1,I_3=0\rangle &=& \frac{1}{\sqrt{2}}  \big(K^{0}\bar{K}^{0}- K^{+} K^{-} \big) \,,
\end{eqnarray}
there could be a mixture of the two resonances and their interference would be responsible for a different  $K^{+} K^{-}$ or $K^{0}\bar{K}^{0}$ production.

It is clear that the three experiments are seeing an $I=1$ resonance in the region $[1700-1800]\mev$ and it is
unlikely that they correspond to three different resonances, but this is something to be settled by further experiments
to which we will come  back. Below this energy there is one $a_0$ resonance very well known, the $a_0(980)$ resonance,
which decays to $\pi\eta$, with an apparent width of about $[50-100]\mev$ \cite{pdg}.

\begin{figure}[h]
\centering
\includegraphics[scale=0.85]{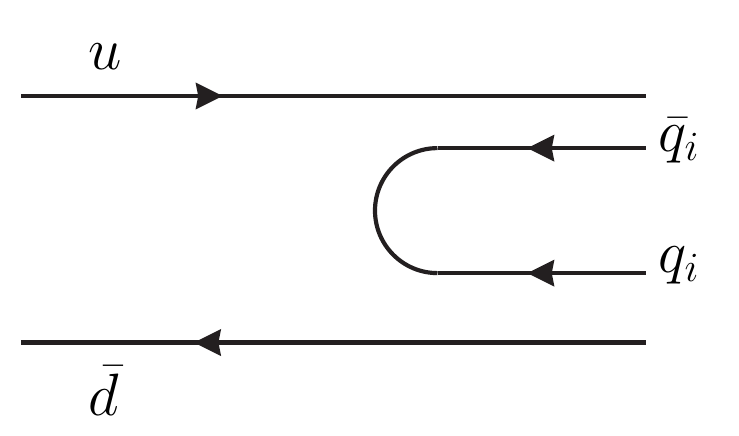}
\caption{Hadronization  of a  $u\bar{d}$ component into two mesons.}
\label{fig:had}
\end{figure}

From the standard $q\bar{q}$ quark model point of view the $a_0^+(980)$ would be $u\bar{d}$. It might be surprising that a state like that
with no strange quarks  decays to $K\bar{K}$. The answer to this lies in the hadronization of the $u\bar{d}$
which gets attached to a $\bar{q}q$ state with the quantum numbers of the vacuum via
\begin{eqnarray}\label{eq:new}
u\bar d \to \sum_i u\, \bar q_i q_i \, \bar d = u( \bar{u}u+\bar{d}d+\bar{s}s)\bar{d}  \,,
\end{eqnarray}
as shown in Fig.~\ref{fig:had}.

It is most convenient to write the $q\bar{q}$ matrix in SU(3) in terms of mesons
given by Eq.~\eqref{eq:p} for pseudoscalar mesons, $P$, and Eq.~\eqref{eq:v} for vector mesons, $V$,
 \begin{eqnarray}\label{eq:p}
P = \left(
  \begin{array}{ccc}
 \frac{\pi^0 }{\sqrt{2}}+ \frac{\eta}{\sqrt{3}}+ \frac{\eta'}{\sqrt{6}}  & \pi^+ & K^+ \\[2mm]
   \pi^- & -\frac{\pi^0 }{\sqrt{2}}+ \frac{\eta}{\sqrt{3}}+ \frac{\eta'}{\sqrt{6}}  & K^0 \\[2mm]
  K^- & \bar{K}^0 & -\frac{\eta}{\sqrt{3}}+\sqrt{\frac{2}{3}}\eta'\\
     \end{array}
    \right)\,,
\end{eqnarray}

\begin{eqnarray}\label{eq:v}
V = \left(
  \begin{array}{ccc}
 \frac{\rho^0 }{\sqrt{2}}+ \frac{\omega}{\sqrt{2}}  & \rho^+ & K^{*+} \\[2mm]
   \rho^- & -\frac{\rho^0 }{\sqrt{2}}+ \frac{\omega}{\sqrt{2}}  &K^{*0} \\[2mm]
  K^{*-} & \bar{K}^{*0} & \phi\\
     \end{array}
    \right)\,.
\end{eqnarray}
We find then for Eq.~\eqref{eq:new}  written in terms of pseudoscalar mesons
\begin{eqnarray} \label{eq:udb1}
u\bar d \to \sum_i  P_{1i} P_{i2} = (P^2)_{12}\,,
\end{eqnarray}
which gives
\begin{eqnarray} \label{eq:udb2}
u\bar d \to  \frac{2}{\sqrt{3}} \eta \pi^+ + K^+  \bar{K}^0
\end{eqnarray}
A rough  estimation  of the decay width of the new $a_0$ with a mass around
$1780\mev$ and the $a_0(980)$, based solely on the phase space and decay in $S$-wave ($\Gamma \sim p_i$, the
momentum of each meson pair in the decay) gives a ratio of 4.27, hence we should expect a width for the new $a_0$
of the order of $[213-427]\mev$. Yet, the width of the new $a_0$  is of the order of $100\mev$, see Eq.~\eqref{eq:a0}.

There seems to be something special in these $a_0$ resonances. Indeed, it was already long ago that the $a_0(980)$ was advocated
as being a $K\bar{K}$ molecule \cite{isgur}. The advent of the chiral unitary approach combining dynamics of chiral Lagrangians with
unitarity in coupled channels \cite{npa} made this idea more quantitative, since the  $a_0(980)$ emerges as a consequence of the interaction
of the coupled channels $\pi\eta$ and $K\bar{K}$.

The success of the chiral unitary approach describing the low-lying scalar mesons $f_0(500)$, $a_0(980)$, $f_0(980)$, $K^{*}_0(700)$, prompted the extension
of these ideas to the interaction of vector mesons in \cite{geng}, using as a source of interaction of the vector mesons the local hidden gauge
approach of \cite{bando}. Interestingly, two resonances were found in the region of energies discussed here, one with $I=0$,  $f_0(1721)$
with $\Gamma \simeq 133\mev$, which was associated to the   $f_0(1710)$   and one with $I=1$,  $a_0(1777)$ with $\Gamma \simeq 148\mev$
(we will call it $a_0(1780)$ in what follows), for which
there was no experimental information at the time of its prediction. The $f_0(1710)$ couples in that approach to
 $K^*\bar{K}^*$, $\rho\rho$, $\omega\omega$,  $\omega\phi$, $\phi\phi$, while the $a_0(1780)$ couples to the channels
 $K^*\bar{K}^*$, $\rho\omega$,  $\rho\phi$, but the largest coupling in both cases is to the $K^*\bar{K}^*$ component, which
 makes the two  resonances qualify roughly as $K^*\bar{K}^*$ molecules in analogy to the $K\bar{K}$
approximate nature of the $a_0(980)$ \cite{npa}. Similar conclusions have been reached more recently in \cite{gulmez}. The smaller binding
of the $a_0(1780)$ comes as a natural consequence of a weaker potential in $I=1$ than in $I=0$ \cite{geng}.

\begin{figure}[h]
\centering
\includegraphics[scale=0.98]{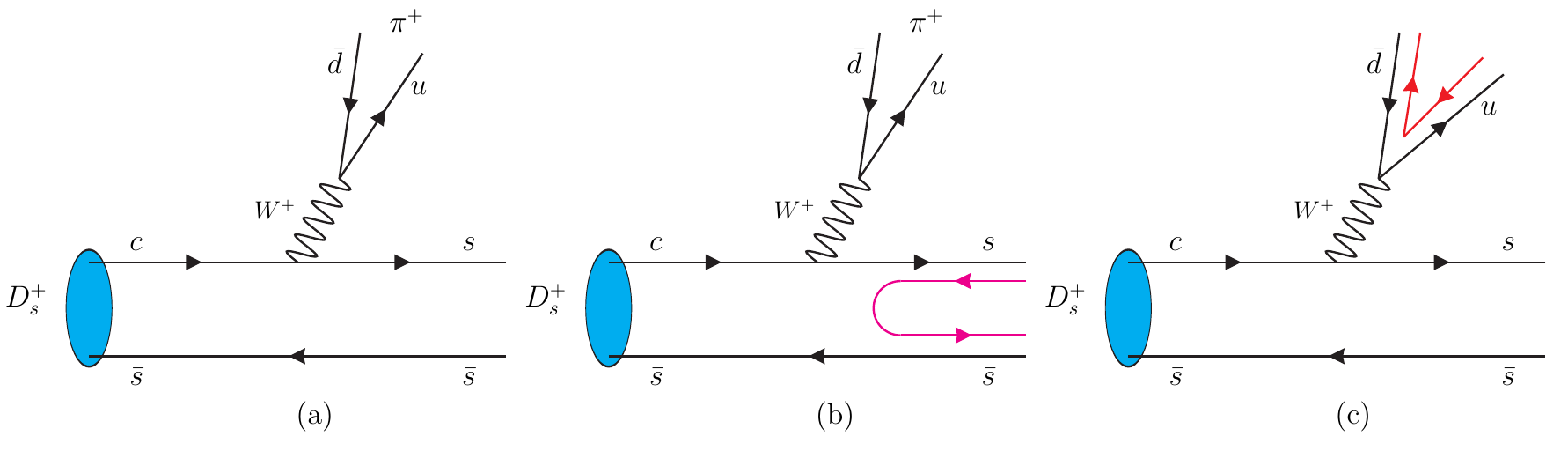}
\caption{Cabibbo-favored decay mode of  $D_s^+$ at the quark level with external emission (a); Hadronization  of the $s\bar{s}$ component (b); Hadronization  of the $\bar{d}u$ component (c).}
\label{fig:abi1}
\end{figure}

While we expect that the new experimental finding of the  $a_0(1817)$ will trigger much theoretical work trying to understand
the nature of the  resonance, it is certainly most appealing to consider that this  resonance  corresponds to the  $a_0(1780)$
 that was predicted in \cite{geng}. The success in other resonances obtained in that approach, which could be
 associated to known states, as well as the success in the predictions of the chiral unitary approach in other sectors,
 make us to consider this as a most likely option. Next step is to test the theory in experiment.
 In this direction, the work done in \cite{daigeng} is illuminating.  The mechanisms for $D_s^+$ decay at the quark level corresponding
 to external emission were considered as shown in Fig.~\ref{fig:abi1}.

\begin{figure}[h]
\centering
\includegraphics[scale=0.88]{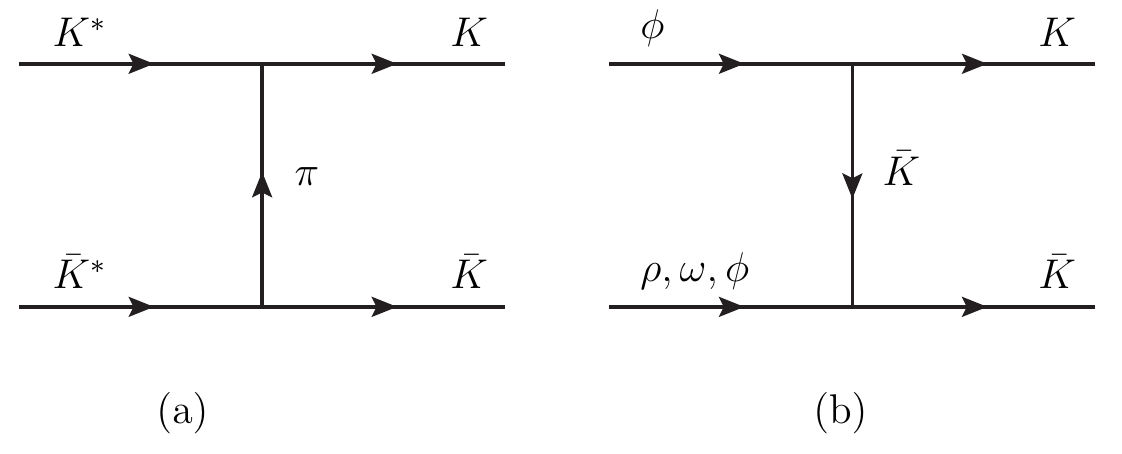}
\caption{$K^* \bar{K}^* \to  K\bar{K}$  transitions driven by $\pi$ exchange (a) and  $\phi (\rho,\omega,\phi) \to K\bar{K}$  transitions driven by $K$ exchange (b).}
\label{fig:tr}
\end{figure}

The hadronization of $s\bar{s}$ or $\bar{d}u$ components was allowed,
following the method of Eqs.~\eqref{eq:udb1},\eqref{eq:udb2},
producing pairs of vector mesons. Subdominant internal emission mechanisms were also considered and the different mechanisms exciting
$f_0(1710)$ and $a_0(1780)$ were identified, and their interference produced a ratio $R_1$ different than unity.
One should mention that while the $f_0(1710)$ and $a_0(1780)$ resonances are made from vector-vector components,
the experiment measures $K\bar{K}$ for which the transition of vector-vector to $K\bar{K}$ must be implemented as depicted in Fig.~\ref{fig:tr}.
Fine tunning two free parameters of the theory, the ratio of Eq.~\eqref{eq:exR1} could be reproduced  in \cite{daigeng}. Yet, the challenge was then
to predict the ratio
\begin{eqnarray}
R_2=\frac{\Gamma(D_s^+ \to \pi^0 a_0(1710)^+ \to \pi^0 K^+  K_S^0)}
{\Gamma(D_s^+ \to \pi^0 ``f_0(1710)" \to  \pi^+ K^+ K^-)}\,,
\end{eqnarray}
where the numerator corresponds to the $\pi^0 a^+_0(1780) $ production, and a value
 \begin{eqnarray}
  R_2^{theo} \simeq 1.31  \pm 0.12 \,\nonumber
\end{eqnarray}
was obtained, from which the branching fraction
\begin{eqnarray}
 {\mathrm{Br}}[D_s^+ \to  \pi^0 a_0(1710)^+; a_0(1710)^+ \to K^+ K_S^0] \simeq  (1.3  \pm 0.4)\times 10^{-3} \,
\end{eqnarray}
resulted.  This was a prediction before this ratio was measured in \cite{beschar}, a fair prediction compared with the branching fraction reported in \cite{beschar} of
$(3.44  \pm 0.52 \pm 0.32)\times 10^{-3}$ considering the smallness of the number for a $D_s$ decay and the amount of
modeling required in \cite{beschar} to extract the $a_0(1817)$ signal. Indeed, in the experiment of \cite{beschar}
(see Fig.2 (a) of  that paper), a prominent peak around $1770\mev$ is seen in the $K_S^0  K^+$ mass distribution,
to which there are five channels contributing, $K^+ \bar{K}^{*0}(892)$, $K_S^0 K^*(892)$, $K^* K^*(1410)^0$,
$a_0(980)^+ \pi^0$, $a_0(1817)^+ \pi^0$, out of which the $a_0(1817)\pi^0 $ is only a small fraction.
On the other hand further developments around the idea of \cite{daigeng} have been carried out very recently including the
$D_s^+ \to K^{*+} \bar{K}^0 \to \pi^+ K_S^0 K_S^0$ mechanism in Ref. \cite{wanggeng}
and the extra one $D_s^+ \to \pi^0 a_0(980) \to \pi^0 K^+ K_S^0$ in \cite{gengxie}, by means of which all the mass distributions in the $D_s^+ \to \pi^+ K_S^0 K_S^0$ and $D_s^+ \to \pi^0 K^+ K_S^0$ reactions are all well reproduced, showing the relevance of the  $a_0(1780)$ state in the process.

It would be most interesting to  devise different reactions where the  $a_0(1817)$ can be
observed and we shall discuss this issue providing some perspective for future experiments.

\begin{itemize}
  \item[1)] In order to clean up the spectrum from unwanted contributions of other resonances, one can make cuts in the mass distributions to eliminate the $K^*(892)$  contributions. This idea has already been used with success in \cite{besraquel} in the
      study of the $D_s^+ \to \pi^+ \pi^0 \eta$ decay, looking at the $a_0(980)$ contribution, in which a cut was performed to eliminate
      the $\eta \rho^+$ contribution by taking $M_{\pi^+ \pi^0}>1\gev$.
  \item[2)] One can investigate strong decays which are more selective concerning isospin conservation than weak decays, where generally
  isospin is not conserved. In this case one can look at different cases. Let us begin by one reaction
  where one would have the violation, $J/\psi \to  \phi K^+ K^- (K^0 \bar{K}^0)$. One can also have $\omega$ instead of $\phi$ in this decay. From the
  perspective of  molecular structure for  $f_0(1710)$ and  $a_0(1780)$ we have to look for combinations of one $\phi$ and the
  vector-vector components. This can be done looking for SU(3) invariants of three vectors, given the fact that $J/\psi$ is an  SU(3) singlet.
  This can be done with the matrix $V$ of Eq.~\eqref{eq:v}, and the invariants are $\langle VVV \rangle$,
  $\langle V\rangle \langle VV \rangle$,  $\langle V\rangle\langle V\rangle\langle V\rangle$, where $\langle\,\rangle$ indicates the trace in  SU(3), but it was shown in \cite{debasliang} and other works that the $\langle VVV \rangle$ structure was prefered by experiments. One finds immediately that the structure coming from  $\langle VVV \rangle$ containing at least one $\phi$ field is
\begin{eqnarray}
3 \,(K^{*+} K^{*-} \, + \, K^{*0}\bar{K}^{*0})\phi +\phi\phi\phi   \,, \nonumber
\end{eqnarray}
which leaves the $K^{*}\bar{K}^{*}$ and $\phi\phi$ channels to produce the  $f_0(1710)$. This
is done through a mechanism shown in Fig.~\ref{fig:4x}.
\begin{figure}[h]
\centering
\includegraphics[scale=0.65]{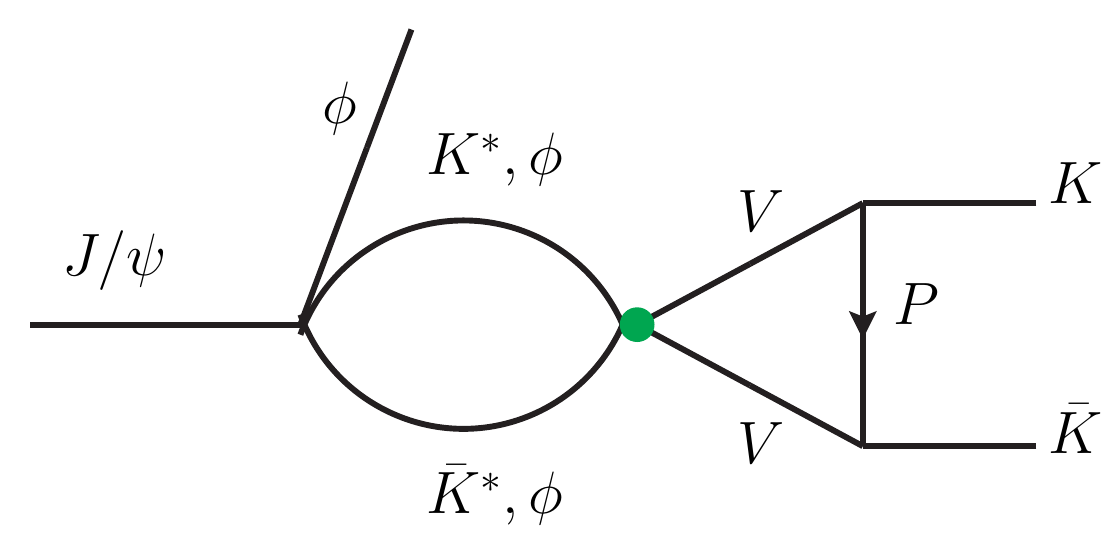}
\caption{Mechanism to produce the $f_0,a_0$  resonances decaying to $K\bar{K}$. The thick dot indicates the transition $T$ matrix $VV \to VV$.}
\label{fig:4x}
\end{figure}
The interesting thing is that given the different masses of $K^{*+}, K^{*-}$ in the loops of Fig.~\ref{fig:4x} and in the construction
of the $VV \to VV$ scattering matrix, isospin will be slightly broken, but sufficiently to allow the production of  $a_0(1780)$
and produce a difference in the $K^+ K^-,K^0 \bar{K}^0$ production in the region of $1780\mev$.
This kind of isospin violation due to different masses of $K^+,K^0$ was already emphasized in \cite{acharov}.
Once again, because of the production of the $a_0(1780)$, we expect to have different $K^+ K^-,K^0 \bar{K}^0$ production rates.

We should note  that because of the width of the $K^{*}$, the effect of the $K^{*+}, K^{*0}$ mass
difference should be much reduced with respect to that in reactions driven by $K^+ K^-$,$K^0 \bar{K}^0$  intermediate states. Yet, through interference effect, and provided one has good statistics, observable effects
are foreseen.

\item[3)] Next one can look at reactions that select  the $I=1$ component in the decay. Let us take for instance
 \begin{eqnarray}
 J/\psi \to  \rho^+ K^0 K^- \,;  \quad   J/\psi \to  \pi^+ K^0 K^-    \,\nonumber
\end{eqnarray}
This will select necessarily the $a_0(1780)^-$. Here we do not have to worry about isospin violation since there is no $f_0(1710)$
with negative charge.  The only problem could stem from a contribution of the $\rho K$ or $\pi K$ interaction. The $\pi K$ should
appear as a strong signal of $K^{*}$, but once again this can be removed eliminating this contribution with cuts in the $\pi K$ invariant
mass. The $\rho K$  would show as a $K_1(1270)$ resonance  peak, but once again one could remove this contribution with cuts.
Alternatively, one can also make the analysis of the reaction with a partial wave analysis, as done in \cite{beschar}.
\item[4)] Finally let us give another suggestion to further pin down the $a_0$ resonance. As discussed above and in  \cite{geng},
the  $a_0(1780)$ resonance couples to $K^{*} \bar{K}^*$, $\rho\omega, \rho\phi$. So far the investigations have been done by looking
at $K\bar{K}$, which comes from the $K^{*} \bar{K}^*$ component. One can look instead to the channels $\rho\omega, \rho\phi$.
Those channels already filter an isospin $I=1$ and would provide a new look at the $a_0(1780) (a_0 (1817))$ resonance.
The $\rho\omega$ channel is open. The $\rho\phi$ channel is closed using the nominal masses of the particles, but
allowed considering the $\rho$ width.
\item[5)] Apart from the former discussion, with suggestion of  new experiments, it is  worth mentioning here some of the
relevant ideas that the discovery of the new $a_0$ state has spurred within different theoretical groups.
In Ref. \cite{xlslz} the new  resonance  has served to  classify existing $a_0$ states into a Regge trajectory, predicting that there should
be a new $a_0$ resonance at $2115\mev$. In Ref. \cite{bingsong} the idea of  \cite{geng} is retaken, showing that the addition of  pseudoscalar-pseudoscalar
channels does not significantly alter the results obtained with just the vector-vector channels.  In that work, a discussion is made about how the partnership of the
known $f_0$  and $a_0$ resonances is affected by the discovery of the new  resonance, which has repercussion on the on-going discussion about the
possibility  of the  $f_0(1500)$ or $f_0(1710)$ states corresponding to  glueball  states, formed purely from gluons and their interaction \cite{amsler}.

\end{itemize}

As we have shown, the new $a_0(1817)$ resonance observed in \cite{beschar} is an important state that sheds light into the structure of scalar mesons in the light quark sector and other relevant issues currently under debate in hadron physics. Given its relevance, the observation of that state in different reactions is very important, as well as the investigation of its decay channels. In this letter we have offered a perspective to make progress in this direction.

\section*{ACKNOWLEDGEMENT}
 This work is partly supported by the National
Natural Science Foundation of China under Grants Nos. 12175066,
11975009, 12147219 and Nos. 11975041, 11735003. This work is also
partly supported by the Spanish Ministerio de Economia y Competitividad (MINECO) and European FEDER funds under Contracts No.
FIS2017-84038-C2-1-P B, PID2020-112777GB-I00, and by Generalitat Valenciana under contract PROMETEO/2020/023. This project has
received funding from the European Union Horizon 2020 research and
innovation programme under the program H2020-INFRAIA-2018-1,
grant agreement No. 824093 of the STRONG-2020 project.

\end{document}